\documentstyle[11pt,newpasp,twoside,epsfig]{article}
\index{Pulsars} \index{Neutron stars}

\def\edcomment#1{\iffalse\marginpar{\raggedright\sl#1\/}\else\relax\fi}
\reversemarginpar

\begin{document}
\title{Testing Pulsar Thermal Evolution Theories with Observation}
\author{Sachiko Tsuruta}
\affil{Montana State University, Bozeman, Montana 59717 USA}

\begin{abstract}
With the successful launch of {\it Chandra} and {\it XMM/Newton}
X-ray space missions combined with the lower-energy band
observations, time has arrived when careful comparison of thermal
evolution theories of isolated neutron stars with observations
will offer a better hope for distinguishing among various
competing neutron star cooling theories. For instance, the latest
theoretical and observational developments may already exclude
both nucleon and kaon direct Urca cooling.  In this way we can now
have a realistic hope for determining various important
properties, such as the composition, superfluidity, the
equation of state and stellar radius. These developments should
help us obtain deeper insight into the properties of dense matter.
\end{abstract}

\section{Introduction}

The launch of the {\it Einstein} Observatory gave the first hope
for detecting thermal radiation directly from the surface of
neutron stars (NSs).  However, the temperatures obtained by the
{\it Einstein} were only the upper limits (e.g., Nomoto \& Tsuruta
1986). {\it ROSAT} offered the first confirmed detections (not
just upper limits) for such surface thermal radiation from at
least three cooling neutron stars, PSR 0656+14, PSR 0630+18
(Geminga) and PSR 1055-52 (e.g., Becker 1995).  Recently the
prospect for measuring the surface temperature of isolated NSs, as
well as obtaining better upper limits, has increased
significantly, thanks to the superior X-ray data from {\it
Chandra} and {\it XMM/Newton}, as well as the data in lower energy
bands from optical-UV telescopes such as {\it Hubble Space
Telescope (HST)}. Consequently, the number of possible surface
temperature detections has already increased to at least nine
(see, e.g., Tsuruta et al. 2002, hereafter T02). Very recently
Chandra offered an important upper limit to PSR J0205+6449 in 3C58
(Slane, Helfand, \& Murray 2002). More detections, as well as
better upper limits, are expected from space missions planned for
the immediate future. These developments have proved to offer
serious `turning points' for the detectability of thermal
radiation directly from the surface of isolated NSs. In addition,
more careful and detailed theoretical investigation of various
input microphysics has already started (see, e.g., Lattimer's
contribution to this volume; Takatsuka et al. 2001; Takatsuka et
al. 2003, hereafter Ta03a; Tamagaki 2003, hereafter Ta03b). In
this report we try to demonstrate that distinguishing among
various competing NS cooling theories has started to become
possible, by careful comparison of improved theories with new
observations (see., e.g., T02; Tsuruta et al. 2003, hereafter
Ts03; Teter et al. 2003, hereafter Te03).

\section{Neutron Star Cooling Theories}

The first detailed cooling calculations (Tsuruta and Cameron 1966)
showed that isolated NSs can be warm enough to be observable as
X-ray sources for about a million years. After a supernova
explosion a newly formed NS first cools via various neutrino
emission mechanisms before the surface photon radiation takes
over. Among the important factors which seriously affect the
nature of NS cooling are: neutrino emission processes,
superfluidity of constituent particles, composition, mass, and the
equation of state (EOS). In this paper, for convenience, the
conventional, slower neutrino cooling mechanisms, such as the
modified Urca, plasmon neutrino and bremsstrahlung processes, will
be called `standard cooling'.  On the other hand, the more
`exotic' extremely fast cooling processes, such as the direct Urca
processes involving nucleons, hyperons, pions, kaons, and quarks,
will be called `nonstandard' processes (see, e.g., Tsuruta 1998,
hereafter T98; Tsuruta \& Teter 2001, hereafter TT01).

The composition of NS interior is predominantly neutrons with only
a small fraction of protons and electrons (and possibly muons
also) when the interior density is not high (the central density
$\rho^c <$ 10$^{15}$ gm/cm$^3$). For higher densities more
`exotic' particles, such as hyperons, pions, kaons and quarks, may
dominate the central core. Therefore, when the star is less
massive and hence less dense, it will be a neutron star with the
interior consisting predominantly of neutrons (no `exotic'
particles) and it will cool with the slower, `standard' neutrino
processes. On the other hand when $\rho^c$ exceeds the transition
density to the exotic matter $\rho_{tr}$, the transition from
nucleons to `exotic' particles takes place. Therefore, more
massive stars, whose $\rho^c$ exceeds $\rho_{tr}$, possesses a
central core consisting of the exotic particles. In that case, the
nonstandard fast cooling takes over\footnote{Note, however, that
if proton fraction in the neutron matter is exceptionally high,
i.e., $> 15\%$, nonstandard fast cooling can take place in a NS
without exotic particles, through the nucleon direct Urca process.
This can happen for a certain type of EOS models which alow such
high proton concentration above a certain critical density
(Lattimer et al. 1991). In order to include this option, in the
following discussion we will call a fast nonstandard process, an
`exotic process', rather than `a process involving exotic
particles'.}.  The observational data suggest that there are at
least two classes of NSs, hotter ones and cooler ones. The most
natural explanation is that hotter stars cool by the standard,
slower cooling processes, while the cooler ones cool by one of the
fast nonstandard processes. This interpretation follows naturally
as the effect of slightly different mass - with hotter stars
somewhat less massive than cooler ones (see Section 3 for the
details).

As the central collapsed star cools after a supernova explosion
and the interior temperature falls below the superfluid critical
temperature, T$^{cr}$, some constituent particles become
superfluid. That causes exponential suppression of both specific
heat (and hence the internal energy) and all neutrino processes
involving these particles.  The net effect is that the star cools
more slowly (and hence the surface temperature and luminosity will
be higher) during the neutrino cooling era due to the suppression
of neutrino emissivity. Therefore, we point out that nonstandard
fast cooling will be no longer so fast if the superfluid energy
gap, which is proportional to T$^{cr}$, is significant.  In
addition to various neutrino cooling mechanisms conventionally
adopted in earlier calculations, recently the `Cooper pair
neutrino emission' (Flowers, Ruderman, \& Sutherland 1976) was
`rediscovered' to be also important under certain circumstances.
This process takes place when the participating particles become
superfluid, and the net effect is to enhance, for some superfluid
models, the neutrino emission right after the superfluidity sets
in.

\section{Most Recent NS Thermal Evolution Models and
Comparison with New Observations}

\subsection{Latest Thermal Evolution Models}
We calculated NS thermal evolution\footnote{We adopt the
expression `thermal evolution' when we include not only cooling
but also heating.} adopting the most up-to-date microphysical
input and a fully general relativistic, `exact' evolutionary code
(i.e., without making isothermal approximations). This code was
originally constructed by Nomoto \& Tsuruta (1987) which has been
continuously up-dated. Our input neutrino emissivity consists of
all possible mechanisms, including Cooper pair emission, both in
the stellar core and crust. The vortex creep heating is also
included, unless otherwise stated.  See Ts03, Te03 for the
details.
The results are summarized in Fig. 1.

Fig. 1a compares thermal evolution of a neutron star and stars
with a pion core.  The EOS adopted is `TNI3U Model' recently
constructed by Ta03a, which is somewhat stiffer than
medium\footnote{Often an EOS is referred to being `stiff' when the
consequent stellar model is more extended and hence less dense,
while it is referred to being `soft' if it is more compact and
denser.}. This EOS refers to pion condensates when density exceeds
$\rho_{tr}^\pi$, the density where transition from neutron matter
to pion matter takes place, which is set to be 4$\rho_0$ (where
$\rho_0$ = 2.8 x 10$^{14}$ gm/cm$^3$ is the nuclear density).
Then, for this particular EOS the central density $\rho^c$ =
$\rho_{tr}^\pi$ for a 1.5M$_\odot$ star\footnote{Note that very
recent observations suggest that mass of an isolated neutron star
may be somewhat higher than 1.4M$_\odot$ (see Section 3.2).}. The
solid curve refers to thermal evolution (including heating) of a
1.4M$_\odot$ star. Since for this star $\rho^c < \rho_{tr}^\pi$,
it consists predominantly of neutrons with only several \%
protons, and it cools by slower `standard' processes. As the core
superfluid model for neutrons we adopt the OPEG-B Model recently
constructed by Ta03b for neutron matter. For proton superfluidity
we adopt the model by Chao, Clark and Yang (1972). The Cooper pair
neutrino emissivity derived by Yakovlev, Levenfish, \& Shibanov
(1999) is adopted for both neutrons and protons in the central
core and neutrons in the inner crustal layers. For the vortex
creep heating we adopt the model with heating parameter K =
10$^{37}$ ergs m$^{-3/2}$ s$^2$ and magnetic field B = 10$^{12}$
Gauss (Umeda, Tsuruta and Nomoto 1994, hereafter UTN94). The long
dashed and dashed curves present thermal evolution of 1.6M$_\odot$
and 1.7M$_\odot$ stars, respectively. These stars cool
predominantly by the nonstandard pion direct Urca process, because
for these more massive stars $\rho^c > \rho_{tr}^\pi$ and hence
the central core consists of pion condensates.  As the superfluid
model for the pion-condensed phase we adopt a medium superfluid
gap model for pion condensates, called the E1-0.6 Model (see Umeda
et al. 1994, hereafter U94;
Takatsuka \& Tamagaki 1982). See Te03 for the details.

\begin{figure}
\begin{center}
\plottwo{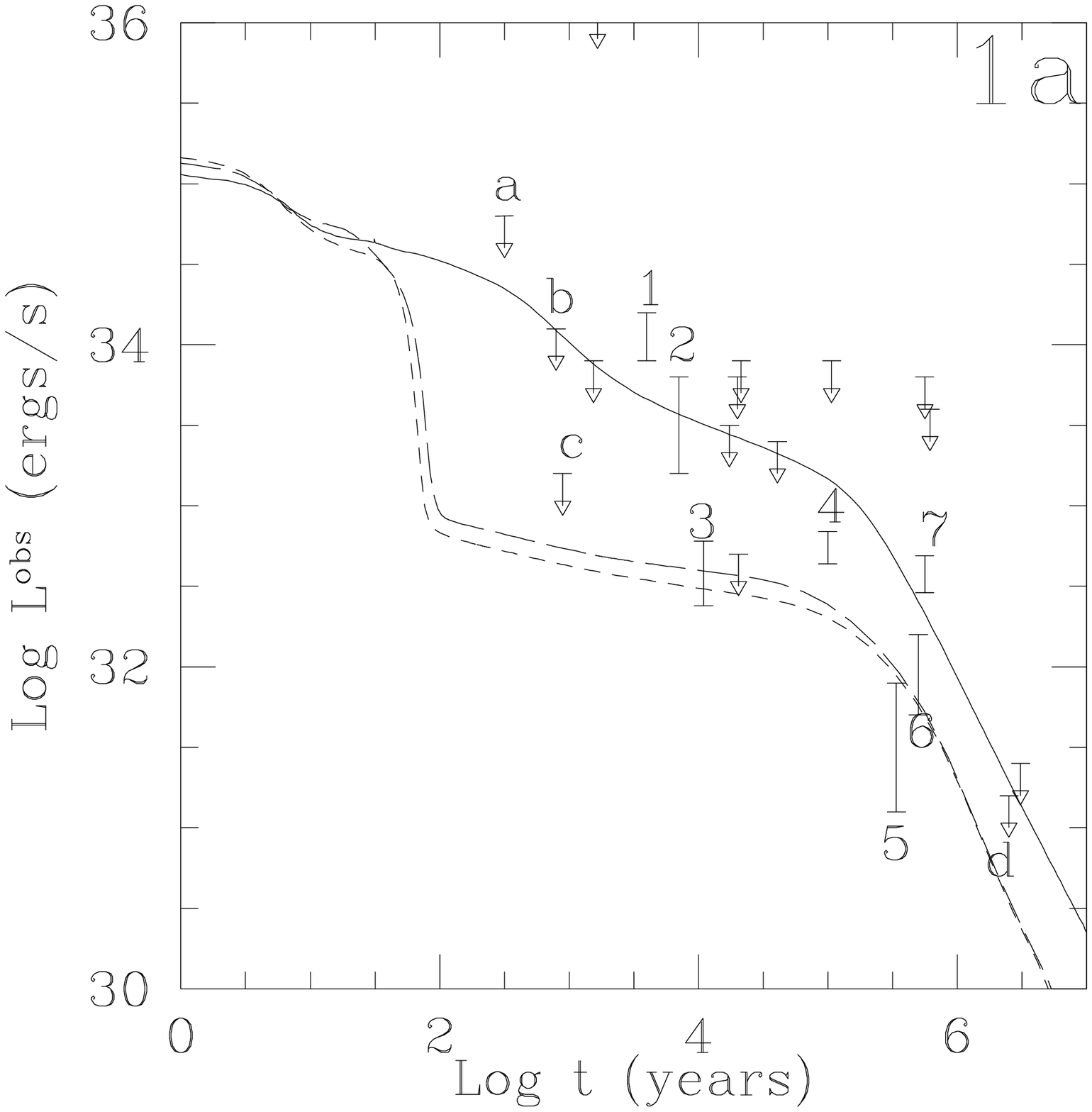}{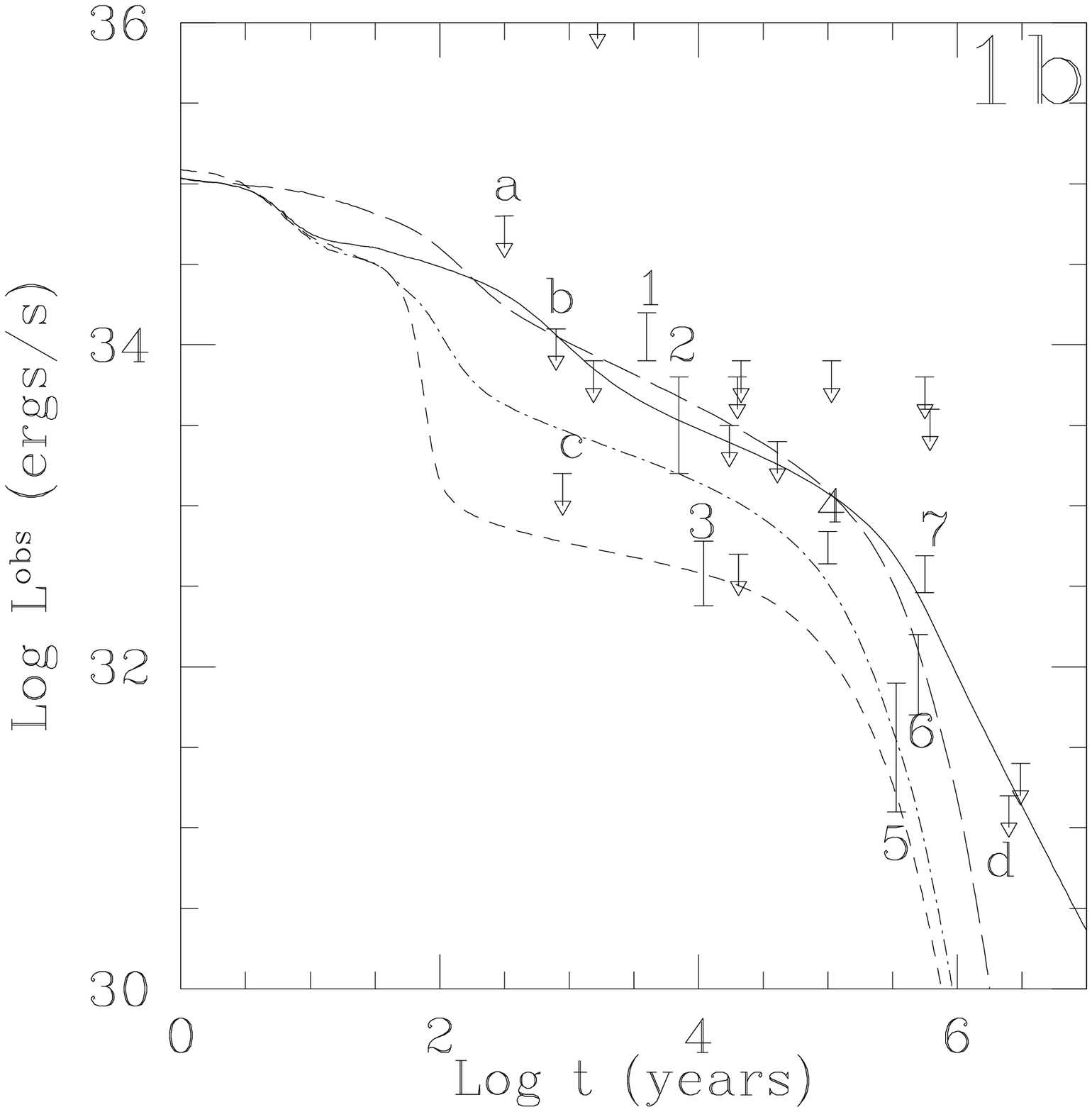}
\caption{
Thermal evolution curves with the newest observational data. The
surface photon luminosity L$^{obs}$ which corresponds to surface
temperature (both to be observed at infinity) is shown as a
function of age. See the text for the details. The vertical bars
refer to confirmed and possible temperature detection data with
error bars, for (1) RX J0822-4300, (2) 1E1207.4-5209, (3) the Vela
pulsar, (4) PSR 0656+14, (5) Geminga, (6) RX J1856-3754, and (7)
PSR 1055-52. (In spite of possible positive detections, sources RX
J0002+62, and RX J0720.4-3125 are not shown because currently
there are still some uncertainties including the age estimate.)
The downward arrows refer to upper limits.  Some interesting
objects are: (a) Cas A point source, (b) Crab pulsar, (c) PSR
J0205+6449 in 3C58, and (d)PSR 1929+10. See Te03 and Ts03 for
references and further details on the data.}
\end{center}
\end{figure}

In Fig. 1b hyperon cooling is compared with standard thermal
evolution curves. The critical transition density from neutron
matter to hyperon matter, $\rho_{tr}^Y$, is set at 4$\rho_0$. For
$\rho > \rho_{tr}^Y$ we adopt the TNI3U EOS for hyperon matter
recently calculated by Ta03a. Other input microphysical parameters
are the same as in Fig. 1a. The solid curve refers to thermal
evolution (heating included) of a 1.4M$_\odot$ NS with the
standard scenario. The dashed curve presents cooling of a
1.6M$_\odot$ hyperon star. For the TNI3U EOS adopted, we find that
$\rho^c$ = $\rho_{tr}^Y$ for a 1.5$_\odot$ star.  Therefore, our
1.6M$_\odot$ star contains a hyperon core and hence the
predominant cooling mechanism is the nonstandard hyperon direct
Urca process. As the superfluid model for hyperons we adopt Ehime
Model for hyperon matter (Ta03a).
See Ts03 for further details.

Fig. 1b also shows two more curves, dot-dashed and long dashed,
for the standard cooling of 1.4M$_\odot$ NSs to explore the effect
of heating and core neutron superfluidity. In the dot-dashed curve
all input parameters are the same as the solid curve except that
heating is switched off. Therefore, by comparing the dot-dashed
curve with the solid curve, we can see the effect of heating,
which we note to be significant. Except that core neutron
superfluidity is not included, the input parameters to the long
dashed curve is the same as the dot-dashed curve.

\subsection{Comparison with Observation}

In Fig. 1 thermal evolution curves are compared with the latest
observational data. The numbers refer to either confirmed or
possible detections while letters refer to some interesting upper
limits. We may note that the data suggest the existence of at
least two classes of sources, hotter stars (e.g., (1) RX
J0822-4300, (2) 1E1207.4-5209 and (7) PSR 1055-52), and cooler
stars (e.g., (c) PSR J0205+6449, (3) the Vela pulsar and (5)
Geminga). The hotter sources are consistent with the solid curves,
the standard cooling of a 1.4M$_\odot$ NS when heating is
included. The source (7) is slightly above the solid curves, but
that is easily explained when the age uncertainty, of at least a
factor of $\sim$ 2 or larger, is taken into account. The source
(1) is somewhat higher than the solid curves, but that is easily
explained by, e.g., the presence of magnetic envelopes with light
elements (e.g., Potenkin et al. 2003).
Comparison of cooler star data with pion and hyperon curves
confirms earlier conclusion (e.g. T98, TT01, T02) that nonstandard
cooling of a more massive star is required for these cooler data.
In that case, we find that significant superfluid suppression is
required if the cooler data are detections. The age uncertainty
should not affect this conclusion, especially for younger cooler
sources such as (c) and (3), because the slope of the curves in
these younger years is small. The data point for (4), PSR 0656+14,
lies between the 1.4 M$_\odot$ and 1.6 M$_\odot$ curves, which
implies its mass is between these masses in our current model.
The conclusion is that all of the observed data are most naturally
and consistently explained as the effects of stellar mass,
superfluidity of the constituent particles, and heating.

At least for binary pulsars, observations offer stringent
constraints on the mass of a NS, to be very close to 1.4M$_\odot$
(e.g., Brown, Weingartner, \& Wijers 1996). If this evidence
extends to isolated NSs also, then the EOS should be such that the
mass of the star whose central density is very close to the
transition density (where the nonstandard process sets in) should
be very close to 1.4M$_\odot$.  With the EOS of medium stiffness
our earlier work (T98, TT01, T02) finds that this transition takes
place for sellar mass near 1.4M$_\odot$.  In the current models
presented here, we take account of the most recent report that
some new observations suggest the mass of isolated NSs to be
somewhat higher (see Nice's contribution to this volume). Then,
the EOS has to be stiffer than medium. This is because a stiffer
EOS corresponds to larger mass for a given central density. That
is why our current model adopts an EOS stiffer than medium. In
conclusion it may be emphasized that in this way {\it comparison
of thermal evolution curves with observed temperature data has a
potential for determining the EOS and hence radius}, if the mass
is fixed. For instance, comparison of this kind may already
eliminate very soft and very stiff EOS.

The qualitative behavior of all nonstandard scenarios is similar
if their transition density is the same (see, e.g., UTN94, T98).
However, here we try to demonstrate that it is still possible to
offer comprehensive assessment of at least which options are more
likely while which are less likely. First of all, we note that all
of the nonstandard mechanisms are too fast to be consistent with
any observed detection data, even with heating (see, e.g., UTN94,
T98). That means significant suppression of neutrino emissivity
due to superfluidity is required, if the cooler data are
detections. However, Takatsuka \& Tamagaki (1997) already showed,
through careful microphysical calculations, that for neutron
matter with such high proton concentration as to permit the
nucleon direct Urca, the superfluid critical temperature T$^{cr}$
should be extremely low, $\sim$ several x 10$^7$ K, not only for
neutrons but also for protons. Here we emphasize that this
conclusion does not depend on the nuclear models adopted for the
calculations. On the other hand, the observed NSs, which are to be
compared with cooling curves, are all hotter (the core temperature
being typically $\sim$ 10$^8$ K to several times 10$^8$ K). That
means {\it the core particles are not yet in the superfluid state}
in these observable NSs. Conclusion is that {\it a star cooling
with nucleon direct Urca would be too cold} to be consistent with
the detection data. The same argument applies to the kaon cooling
also (Takatsuka \& Tamagaki 1995). Further details are found in
TT01, T02, Ts03, Te03. As to the hyperon option Fig. 1b shows that
it will be a viable option if Ehime Model adopted is valid.
However, recently the Gifu-Kyoto nuclear experimental group
(Takahashi et al. 2001) reports that the superfluid gap for
hyperons would be much smaller. If so, hyperon cooling also would
be in trouble with the same reason as for nucleon direct URCA if
the cooler data are detections, due to lack of superfluid
suppression. The problem for quarks is that theoretically there
are still too many unknown factors to offer the level of
exploration possible for the other options (see Ts03). On the
other hand, detailed theoretical investigations have already shown
that T$^{cr}$ for pion condensates is realistically high and hence
pions will safely be in a superfluid state.
The conclusion is that pion cooling is consistent with both theory
and observation. See T98, TT01, T02, Ts03, and Te03 for further details.

\section{NS Cooling Models by Other Groups}

Various other groups have calculated neutron star thermal
evolution. A comprehensive review is found in, e.g., T98. Due to
lack of space here we comment on only the latest work by
Yakovlev's group, e.g., Yakovlev et al. 2003. Although very often
these authors adopted simplified `toy models' with the isothermal
and other various approximations, their results and ours generally
agree, at least qualitatively, when similar input is
applied\footnote{Yakovlev and Haensel (2002) state that the
observation data of T02 are wrong. However, this is due to their
mistake in the business of correctly converting surface
temperature to luminosity. Also they misunderstood some of the
models presented in T02. See Ts03 for the details.}. There are,
however, some serious differences in our interpretation of the
results. For instance,
(i) since in the presence of core superfluid neutrons standard
cooling is not hot enough for the data of PSR 1055, these authors
conclude that neutron superfluidity must be so weak as to be
negligible. However, this conclusion contradicts with the results
of serious microphysical calculations of neutron superfluidity
(see Ta02b, T02, Ts03 for the details), which find that neutron
superfluidity should not be so weak for normal neutron matter with
small proton concentration where the standard cooling operates. On
the other hand, we have shown (see Fig. 1) that this discrepancy
disappears when heating is included properly in calculations of
thermal evolution of isolated (not binary) neutron stars - and
hence no contradiction with theories.
(ii) For detailed comparison of nonstandard cooling with
the cooler star data, these authors chose the nucleon direct Urca
process, which they called Durca. To be consistent with the
apparent need for no (or only very weak) neutron superfluidity for
slower standard cooling to explain the hot data, these authors
conclude that observational data require strong proton
superfluidity to supply sufficient superfluid suppression, in
order to explain the cooler detection data by Durca. However, it
was already emphasized in Section 3.2 that theoretically, proton
superfluidity must be so weak that its suppression effect should
be negligible for models which will allow Durca.  These authors
adopted proton superfluid model for normal neutron matter with
small proton fraction, which fails to apply when the proton
concentration is high enough so that Durca can operate.

\section{Concluding Remarks}

We have shown that the most up-to-date observed temperature data
are consistent with the current thermal evolution theories of
isolated NSs if less massive stars cool by standard cooling while
more massive stars cool with nonstandard cooling, and if heating
is also in operation. Among various nonstandard cooling scenarios,
both Durca and kaon cooling should be excluded if the cooler data
are detections. The major reason is that for Durca to be
operative, high proton concentration is required, which weakens
superfluidity of both protons and neutrons. Similar argument
applies to kaon cooling. Hyperon cooling may be in trouble if the
cooler data are detections and if the hyperon superfluid gap
should be so small as reported by recent nuclear experiments.  On
the other hand, pion cooling is still consistent with both
observation and theory.  The important conclusion is that if the
cooler data are detections, {\it presence of `exotic' particles,
most likely pion condensates, will be required within a very dense
star}.

The capability of constraining the composition of NS interior
matter purely through observation alone will be limited, and hence
it will be very important to {\it exhaust all theoretical
resources.}  Theoretical uncertainties are also very large,
especially in the supranuclear density regime.  However, here we
emphasize that we should still be able to set {\it acceptable
ranges} of theoretical feasibility, at least to separate models
more-likely from those less-likely.  More and better data expected
soon from {\it Chandra}, {\it XMM/Newton}, {\it HST}, and future
third generation missions, when combined with improved theories,
should give still better insight to some fundamental problems in
dense matter physics.

\acknowledgments We acknowledge with special thanks contributions
by our collaborators, W. Chandler, M.A. Teter, T. Takatsuka, R.
Tamagaki, K. Fukumura, G. Pavlov, K. Nomoto, T. Tatsumi, and H.
Umeda, to the results presented in this paper.
Our work for this paper has been supported in part by NASA grants
NAG5-3159, NAG5-12079, AR3-4004A, and G02-3097X.

\end{document}